# Multicasting over Overlay Networks – A Critical Review

M.F.M Firdhous
Faculty of Information Technology,
University of Moratuwa,
Moratuwa,
Sri Lanka.
Mohamed.Firdhous@uom.lk

*Abstract* – **Multicasting technology uses the minimum network resources to serve multiple clients by duplicating the data packets at the closest possible point to the clients. This way at most only one data packets travels down a network link at any one time irrespective of how many clients receive this packet. Traditionally multicasting has been implemented over a specialized network built using multicast routers. This kind of network has the drawback of requiring the deployment of special routers that are more expensive than ordinary routers. Recently there is new interest in delivering multicast traffic over application layer overlay networks. Application layer overlay networks though built on top of the physical network, behave like an independent virtual network made up of only logical links between the nodes. Several authors have proposed systems, mechanisms and protocols for the implementation of multicast media streaming over overlay networks. In this paper, the author takes a critical look at these systems and mechanism with special reference to their strengths and weaknesses.**

## I. INTRODUCTION

Media multicasting is one of the most attractive applications that can exploit the network resources least while delivering the most to the clients. Multicast is a very efficient technology that can be used to deliver the same content to multiple clients simultaneously with minimum bandwidth and server loading. The use of this technology is many and web TV, IP TV, web radio, online delivery of teaching are few of them [1-4]. In all these applications, the server continues to deliver the content while clients can join and leave the network any time to receive the content, but they would receive it only from where they joined the stream.

Traditionally multicasting was tied to the underlying network with special multicast routers making the necessary backbone. Multicast routers are special type of routers with the capability of duplicating and delivering the same data packet to many outgoing links depending on which links clients reside downstream. Traditional IP based routers are unicast that receive data packets on one link and either forward that packet only to one outgoing link or drop it depending on the routing table entries. Broadcasting is totally disabled in the internet due to the unnecessary congestion caused by broadcast traffic that may bring the entire internet down in a short time.

Using the network layer multicast routers to create the backbone of the internet is not that attractive as these routers are more expensive compared to unicast routers. Also, the absence of a multicast router at any point in the internet would defeat the objective of multicasting throughout the internet downstream from that point. Chu et al., have proposed to replace the multicast routers with peer to peer clients for duplicating and forwarding the packets to downstream clients [5]. In this arrangement, the duplicating and forwarding operation that makes multicasting attractive compared to unicast and broadcast would be carried out at the application layer. The multimedia application installed in end nodes would carry out the duplicating and forwarding operation in addition to the display of the content to the downstream nodes that request the stream from an upstream node.

A peer to peer network is formed by nodes of equal capability and act both as client and server at the same time depending on the function performed. Peer to peer networks and applications have advantages over traditional client server, such as the elimination of single point of failure, balance the network load uniformly and to provide alternate path routing easily in case of link failures [6]. A peer to peer network forms an overlay network on top of the existing network infrastructure. The formation of this overlay network by the peer to peer nodes make the resulting network resilient to changes in the underlying network such as router and link failures and congestion.

An overlay network is a computer network which is built on top of another network. Nodes in the overlay can be thought of as being connected by virtual or logical links, each of which corresponds to a path, perhaps through many physical links, in the underlying network. Usually overlay network run at the application layer of the TCP/IP stack making use of the underlying layer and independent of them [7]. Figure 1 shows the basic architecture of an overlay network. The nodes in an overlay network will form a network architecture of their own that may be totally different and independent of the underlying network.

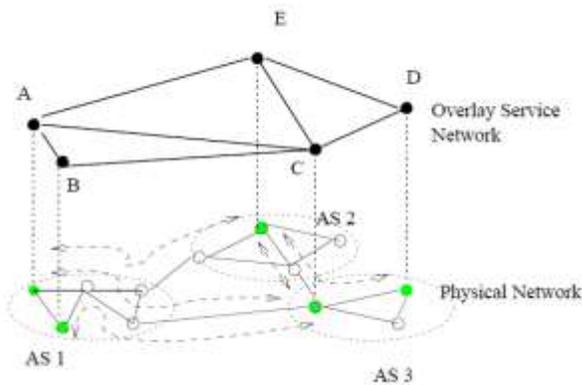

Figure 1: An Overlay Network

## II. MULTICASTING OVER OVERLAY NETWORKS

Several authors have proposed protocols and services how to implement overlay based multicast over the internet. In this section, an in depth analysis would be carried out on the some of the most recent protocols and application in terms of their architecture, advantages and disadvantages.

Chen et al., have proposed ACOM Any-source Capacity-constrained Overlay Multicast system. This system is made up of three multicast algorithms namely Random Walk, Limited Flooding, and Probabilistic Flooding on top of a non-DHT (Distributed Hash table) overlay network with simple structures [8]. ACOM divides the receiving nodes into multicast groups based on the upload bandwidth of a node as the upload bandwidth is determining factor as a node may be required to transfer multiple copies of the same packet over the uplink. The number of nodes in a multicast group hence depends on the capacity of the uplink making the nodes with higher capacity to support large number of neighbors. An overlay network is established for each multicast group that transforms the multicast stream to a broadcast stream with the scope of the overlay.

The overlay network is made up of two components, namely an unrestricted ring that is fundamentally different from the location specific DHT based ring and the random graph among the nodes. The ring maintenance is carried out by requiring the nodes to the next node as a successor and a few other nodes in order to avoid the problem of ring breakage due to node leaving the network.

Packet delivery with the overlay network is carried out using a two phase process. In Phase 1, packets are forwarded to a random number of neighbors within a specific number (k – a system parameter) of hops. This follows a tree structure of delivery in Phase 1. For the purpose of delivery in Phase 2, the ring is partitioned into segments and each Phase 1 node is made responsible for a segment and required to deliver the packet to its successor which in turn forwards to its successor until the packet reaches a node that has already received the packet.

In practice what ACOM does is to forward the packet in Phase 1 using a tree structure to a number of random nodes and then in Phase 2 the packet is forwarded in a unicast fashion down the network until all the nodes within the multicast subgroup receives the packet.

Even though ACOM presents several advantages such as maintenance of virtual tree in place of multicast tree bound to the physical networks, it has certain disadvantages too. The main disadvantage is the total disregard of the physical distances when setting up of the virtual tree. ACOM also has certain other disadvantages in the formation and maintenance of the ring network.

Liu and Ma have proposed a framework called Hierarchy Overlay Multicast Network based on Transport Layer Multi-homing (HOMN-SCTP) [9]. This framework uses transport layer multi-homing techniques. HOMN-SCTP can be shared by a variety of applications, and provide scalable, efficient, and practical multicast support for a variety of group communication applications. The main component of the framework is the Service Broker (SvB) that is made up of the cores of of HOMN-SCTP and Bandwidth- Satisfied Overlay Multicast (BSOM). The BSOM searches for multicast paths to form overlay networks for upper layer QoS-sensitive applications, and balance overlay traffic load on SvBs and overlay links.

HOMN-SCTP is capable of supporting multiple applications on it and helps these applications meet the required QoS requirements. Since it is built on top of TCP, error control is totally delegated to the underlying layer. Nevertheless HOMN-SCTP is not suitable for applications such as streaming media that run on top of UDP.

Fair Load Sharing Scheme proposed by Makishi et al., concentrates on improving throughput instead of reducing the delay on multisource multimedia delivery such as video conferencing using Application Layer Multicasting (ALM) [10]. This scheme is based on tree structure and tries to find the tree with the highest possible throughput. The main objective of the protocol is to improve the receiving bit rate of the node that is having lowest bit rate out of all receivers. The proposed scheme is an autonomous distributed protocol that constructs the ALM where each node refines its own subtree. The end result of this refinement of subtrees is the automatic refinement of the entire tree structure.

This scheme assumes that there is backbone ring network that has unlimited bandwidth and all the receiving nodes can connect to this ring. The links from the backbone to the end nodes make the tree network which this protocol tries to improve for the receiver bandwidth. The end nodes initially make a basic (ad-hoc) tree network that would be improved iteratively over time until the all the nodes receive an optimum bandwidth in terms of bit rate. Once the basic tree structure has been established using unicast links, the nodes communicate with each other exchanging estimation queries with communication quality in terms of bandwidth to be allocated and the delay. When a better path than that is currently being used is found, the tree readjusts itself moving to the node with the better quality. This process is continued until the optimum link allocation is achieved.

Peng and Zhang have discussed the problem associated with the intranet based multimedia education platform [4]. They mainly analyze the principles of this education platform, but in the process they introduce Computer Assisted Instruction (CAI) multicasting network as the backbone of the network. The CAI multicasting network is based on the IPv4 special class D multicasting (224.0.0.0 ~ 239.255.255.255) address block. The multicasting network has been implemented using the Winsock2 mechanism. Once the server node (teacher) has initiated the stream the user (students) can join the multicast network and receive the multicast stream on their computers. This system is suitable only for in class teaching or within limited distance where high networking resources are available.

Reduced delay and delay jitter are very important to viewing quality of any video from the viewers' point. In traditional IP multicasting RSVP and DiffServ algorithms were used to reserve resources prior to starting the streaming. Each router on the path from the server to the client uses a dynamic scheduling algorithm to deliver the packet based on the QoS requirements. Szymanski and Gilbert have proposed a Guaranteed Rate (GR) scheduling algorithm for computing the transmission schedules for each IQ packet-switched IP router in a multicast tree. The simulation results have shown that this algorithm manages on average two packets in a queue resulting in very low delay and jitter which can practically ignored as zero jitter [11].

Even though this algorithm virtually eliminates delay jitter, it is bound to the underlying layers as it needs to be implemented on routers.

Lua et al., have proposed a Network Aware Geometric Overlay Multicast Streaming Network. This network exploits the locality of the nodes in the underlay for the purpose of node placement, routing and multicasting. This protocol divides the nodes into two groups called SuperPeers and Peers. SuperPeers form the low latency, high bandwidth backbone and the Peer connect to the nearest SuperPeer to receive the content [12].

SuperPeers have been elected based on two criteria, namely: the SuperPeer should have sufficient resources to serve other SuperPeers and Peers and they must be reliable in terms of stability not join and leave the network very frequently. Network embedding algorithm computes node coordinates and geometric distances between nodes to estimate the performance metrics of the underlying network such as latency. Peers joining the overlay network calculate the total Round Trip Time (RTT) to the SuperPeers and join the SuperPeer that has the lowest RTT. A SuperPeer joining the network calculates the RTT to all the existing SuperPeers and joins the ones with the lowest RTT and then creates connection with other six SuperPeers around it.

When a SuperPeer leaves the network, the other SuperPeers would detect this by the loss of heartbeat signal from the node that has left and will reorganize themselves by sending discovery broadcast messages to all the SuperPeers. The Peers who are affected by the leaving of a SuperPeer need to reconnect to the overlay network by selecting the nearest SuperPeer.

The main strengths of this scheme can be summarized as it has good performance in terms of latency and efficient transmission of packets via the high speed backbone network whereas the weaknesses include the heavy dependence on the geographical locality, election of SuperPeers.

Pompili et al., have presented two algorithms called DIfferentiated service Multicast algorithm for Internet Resource Optimization (DIMRO) and DIfferentiated service Multicast algorithm for Internet Resource Optimization in Groupshared applications (DIMRO-GS) to build virtual multicast trees on an overlay network [13].

DIMRO constructs virtual source-rooted multicast trees for source-specific applications taking the virtual link available bandwidth into account. This avoids traffic congestion and fluctuation on the underlay network. Traffic congestion in the underlay would cause low performance. This keeps the average link utilization low by distributing data flows among the least loaded links. (DIMRO-GS) builds a virtual shared tree for group-shared applications by connecting each member node to all the other member nodes with a source-rooted tree computed using DIMRO.

Both these algorithms support service differentiation without the support of the underlying layers. Applications with less stringent QoS requirements reuse resources already exploited by members with more stringent requirements. Better utilization of network bandwidth and improved QoS are achieved due to this service differentiation.

System built using these algorithms would result in better performance due to differentiation of applications based on QoS requirements, but node dynamic may bring the quality of the system down.

Wang et al., have proposed an adapted routing scheme that minimizes delay and bandwidth consumption [14]. This routing algorithm creates an optimum balanced tree where the classical Dijkstra's algorithm is used to compute the shortest path between two nodes. In this scheme the Optimal Balance of Delay and Bandwidth consumption (OBDB) is formulated as:

$$OBDB = (1-\alpha)D + \alpha B$$

Where D is the minimal delay criteria and B is the minimal bandwidth consumption criteria.

From the above formula, it can be seen that one routing method may lead to another bad performance. The value of $\alpha$ is calculated based on the nature of the application. As the delay and bandwidth can be tuned to suit application requirements, applications performance can be controlled. But these algorithms will have performance issues in the face of node dynamics.

Kaafar et al., have proposed an overlay multicast tree construction scheme called LCC: Locate, Cluster and Conquer. The objective of this algorithm is to address scalability and efficiency issues [15]. The scheme is made up of two phases. One is a selective locating phase and the other one is the overlay construction phase. The selective locating phase algorithm locates the closest existing set of nodes (cluster) in the overlay for a newcomer. The algorithm does not need full knowledge of the network to carry out this operation, partial knowledge of location-information of participating nodes is sufficient for this operation. It then allows avoiding initially randomly-connected structures with neither virtual coordinates system embedding nor fixed landmarks measurements. Then, on the basis of this locating process, the overlay construction phase consists in building and managing a topology-aware clustered hierarchical overlay.

This algorithm builds an efficient initial tree architecture with partial knowledge of the network but node dynamics may result in poor performance with time.

Walters et al., have studied the effect of the attack by adversaries after they become members of the overlay network [16]. Most of the overlay protocol can handle benign node failures and recover from those failures with relative ease, but they all fail when adversaries in the network start attacking the nodes. In this study, they have identified, demonstrated and mitigated insider attacks against measurement-based adaptation mechanisms in unstructured multicast overlay networks.

Attacks usually target the overlay network construction, maintenance, and availability and allow malicious nodes to control significant traffic in the network, facilitating selective forwarding, traffic analysis, and overlay partitioning. The techniques proposed in this work decrease the number of incorrect or unnecessary adaptations by using outlier detection. The proposed solution is based on the performance of spatial and temporal outlier analysis on measured and probed metrics to allow an honest node to make better use of available information before making an adaptation decision.

This algorithm creates a resilient overlay network in the both structured and unstructured overlay network in the presence of malicious attacks. But the strict nature of the algorithm may delay the adaptation of the network in the event of node dynamics disrupting the flow of information.

Alipour et al., have proposed an overlay protocol known as Multicast Tree Protocol (OMTP) [17]. This protocol can be used to build an overlay tree that reduces the latency between any two pair of nodes. The delay between the nodes has been reduced by adding a shortcut link by calculating the utility link between two groups.

The main advantage of this algorithm is the efficient data transfer but the efficiency may be affected by node dynamics in the overlay network.

Bista has proposed a protocol where the nodes informs the other nodes its leaving time when it joins the network [18]. Using this leaving time information new nodes are joined at the tree in such a manner early leaving nodes would make the leaf nodes down the line and the nodes that would stay longer would be at the higher levels. It has also been proposed a proactive recovery mechanism so that even if an upstream node leaves the tree, the downstream nodes can rejoin at predetermined nodes immediately, so that the recovery time of the disrupted nodes is the minimum.

These algorithms have several drawbacks including, the prior notice of the duration of stay in the network, arrangement of nodes based on the time of stay and central control to manage the node information.

Gao et al., have proposed a hybrid network combining the IP multicast network and the mesh overlay network [19]. The main objectives of the design were to build a network with high performance (low end-to-end transmission latency, high bandwidth mesh overlay links), low end-to-end hop count and high reliability.

This algorithm results in a good structure combining multicast and overlay, but the resulting network is still dependant on the physical network and managing the mesh network is expensive in terms of network resources.

Wang et al., have proposed hybrid overlay network combining a tree and mesh networks called mTreebone [20]. In the mTreebone, the tree forms the backbone and local nodes make a mesh to share content. The backbone tree network has been constructed by identifying the stable nodes as the churn of the backbone would be more expensive in terms of service disruption than the leaf node churn. On top of the tree based overlay network a mesh network has been created in order to handle the effect of node dynamics. Since the mesh network has been updated regularly using keep alive packets, any change in the mesh network immediately notified to all the other mesh nodes. In this design heuristics have been used to predict the stability of the nodes assuming the age of the in the network to be directly proportional to the probability of it staying longer in the network. That is longer a node stays in the network, larger the probability it would stay even further and more stable.

This algorithm results in a resilient overlay network but the network may carry duplicate packets in some part of the network. Also, the maintenance of the mesh network requires large network resources.

Guo and Jha have shown that the main problem in overlay based multicast networks is to optimize routing among CDN servers in the multicast overlay backbone in such a manner that it reduces the maximal end-to-end latency from the origin server to all end hosts is minimized [21]. They have identified this as the Host-Aware Routing Problem (HARP) in the multicast overlay backbone. The main reason for HARP is the last mile latency between end hosts and their corresponding proxy servers. The author of this paper have framed HARP as a constrained spanning tree problem and shown that it is NP-hard. As a solution they have presented a distributed algorithm for HARP. They also have provided a genetic algorithm (GA) to validate the quality of the distributed algorithm.

This structure results in a low latency routing path improving QoS but the non-consideration of node dynamics will affect the overall quality of the network.

Table I summarize the systems, algorithms and mechanism discussed above with special reference to their advantages and disadvantages. hows the comparison of these works with special reference to their advantages and disadvantages.

TABLE I: COMPARISON OF MULTICASTING OVERLAY NETWORKS

|  | Work | System/Protocol Proposed | Advantages | Disadvantages |
|---|---|---|---|---|
| 1. | [4] | Computer Assisted Instruction (CAI) Multicasting Network | This system is suitable for in class teaching or teaching within a limited area using the high technology to larger classes. | This system uses the existing technologies to build a platform and hence no new technology has been introduced in terms of multicasting.<br><br>This technique may work well on an intranet but cannot be ported to the internet as the internet lacks support for class D multicast addresses. |
| 2. | [8] | Any-source Capacity-constrained Overlay Multicast System | Only maintenance of virtual tree is necessary, no strict multicast trees are maintained.<br><br>Resilient to node dynamics as node maintain multiple neighbors in its neighbor table.<br><br>Simple maintenance of unrestricted ring. | The random nature of the tree formation in Phase 1 totally ignores the physical distance from the source to that node. The researchers have considered this as an advantage, but this will make unnecessary delay in delivering a packet to a node that may be physically closer to the source but not selected as a Phase 1 node.<br><br>Even though, the unrestricted ring is better than location specific DHT ring in terms of node creation and maintenance, the unrestricted node is inefficient in forwarding packets and the logical neighbors may not always be neighbors physically.<br><br>Phase 2 forwarding is essentially unicast and cascaded in ACOM which may create a lot of delay in the ultimate delivery to the last node. The problem will be more severe in case of a high capacity Phase 1 node as it will have a large number of neighbors.<br><br>Converting any portion of the multicast network to a broadcast network is inherently in efficient as broadcast is an inefficient protocol in terms of network utilization. |
| 3. | [9] | Hierarchy Overlay Multicast Network | This framework builds an infrastructure on which multiple applications can be built.<br><br>The framework helps application to meet the required QoS.<br><br>The framework makes use of the facilities of the underlying layer for the error control as it is built on top of TCP. | The framework has been built on top of TCP and hence is not suitable for best effort services running on UDP especially streaming media applications.<br><br>TCP is a high overhead protocol compared to UDP and hence, this protocol would inherently have high overhead.<br><br>This framework will not meet the QoS requirement in terms of delay and delay jitter on high loss links as TCP would create delay and delay jitter on high loss networks and hence not suitable for real time applications like media streaming. |
| 4. | [10] | Fair Load Sharing Scheme | This scheme works purely on the application layer without directly depending on the lower layers.<br><br>The protocol results in the optimum receiving tree structure for a given situation. | This protocol is only suitable for network with stable receiving nodes.<br><br>Dynamic nature of the nodes joining and leaving the network will make the tree fail as there is no mechanism in the protocol to handle node dynamics.<br><br>The dynamic nature of the internet would make the tree structure oscillating as the communication quality heavily depends on the dynamics of the underlying network.<br><br>Tree management is costly in terms of information exchanged between the nodes as the continuous information exchange is needed. |
| 5. | [11] | Guaranteed Rate (GR) Scheduling Algorithm | This algorithm virtually eliminates delay jitter in received video stream resulting in high quality reception.<br><br>The algorithm is simple enough to implement on any IP router. | This algorithm needs to be implemented on IP routers and hence bound to the underlay network.<br><br>This does not provide a solution to the existing problem of how to use the existing network as it is to transport multimedia streaming without depending on the underlay network. |
| 6. | [12] | Network Aware Geometric Overlay Multicast Streaming Network | This algorithm has good performance in terms of latency as the physical location of the node has been computed and used as a parameter in forming the overlay network.<br><br>Breaking the network into two layers result in efficient transmission of | Since the algorithm heavily depends on the geographical location, an efficient geographical location computing algorithm is vital in addition to managing the overlay network.<br><br>Since the election of SuperPeers depends on the condition that the SuperPeer should have high dependability in terms of churn frequency, past historical data may be necessary to |

| | | | packets as the SuperPeers are connected to each other via a high speed backbone network. | determine the reliability of a node accurately. SuperPeers should also have sufficient resources to support other SuperPeers and Peers in the network, if the resource availability information is obtained from the node itself, this would be an invitation to rogue nodes to highjack the entire backbone network using false information. The Peers that are affected by a SuperPeer leaving the network do not have automatic transfer to another SuperPeer, this drastically affect the reliability of the entire multicast operation. |
|---|---|---|---|---|
| 7. | [13] | DIfferentiated service Multicast algorithm for Internet Resource Optimization (DIMRO) and DIfferentiated service Multicast algorithm for Internet Resource Optimization in Groupshared applications (DIMRO-GS) | These algorithms result in good performance for both QoS stringent applications and non QoS stringent applications due to service differentiation. | The node dynamics has not been considered when designing these protocols, hence node dynamics would drastically bring the quality of the network down. |
| 8. | [14] | Adapted Routing Scheme | These algorithms results in better performance for any kind of application as delay and bandwidth can be tuned to meet the application requirements. | The node dynamics has not been considered in this routing scheme and hence node dynamics would drastically bring the quality of the network down. |
| 9. | [15] | LCC : Locate, Cluster and Conquer multicast tree | This algorithm creates the initial tree architecture very efficiently with partial knowledge of the overlay network. | The node dynamics has not been considered in this routing scheme and hence node dynamics would drastically bring the quality of the network down. |
| 10. | [16] | | This algorithm creates a resilient overlay network in the face of attacks by malicious nodes. The algorithm works on unstructured overlays and hence can be easily adapted to structured overlay networks too. | The strict nature of the algorithm delays the adaptation of the network to genuine node dynamics. The delay in network adaptation may result in unnecessary disruptions to the streaming and affect the quality of service of the application. |
| 11. | [17] | Multicast Tree Protocol | This algorithm results in efficient data transfer between the source to destination in terms of reduced delay. | The node dynamics has not been considered in constructing the tree and hence node churn would result in broken trees affecting the downstream nodes. |
| 12. | [18] | Proactive Fault Resilient Overlay Multicast Network | A proactive mechanism has been proposed that keeps the downstream nodes informed when to expect an upstream node leaving for the purpose of reconnection. | The setup shown in this algorithm is very artificial as the nodes need to inform the other nodes their duration of stay at the beginning itself. This against the basic spirit of the overlay network where nodes can join and leave the network at their choice. Arranging the nodes in the in the reverse order of the duration of the stay is very impractical as it may result in an inefficient structure if the longest staying node is very far away from the source. Node dynamics has not been properly considered in this design as the affected nodes need to rejoin the network themselves. A central control would be needed to keep the information about the nodes joining time and duration of stay for proper operation of these protocols. |
| 13. | [19] | Hybrid IP multicast mesh overlay network | Results in a good structure combining both IP multicast and mesh overlay. | This network is not totally independent of the underlying network as it depends on the IP multicast protocol. Establishing a full mesh network is not efficient as it would need a large memory maintain information about each and every node. This problem would become more acute as the network size grows to very large. This would result in scalability problems. Maintaining the full mesh information also has the problem of updating the mesh information table as all the nodes need to update the information table every time a node |

| | | | | joins or leaves the network. This would result in instability in the network as the network grows. |
|---|---|---|---|---|
| 14. | [20] | mTreebone | This structure show better resilience to node dynamics compared to pure tree structured overlay networks. Also this structure would have lesser load on the backbone tree as it would carry only one data packet at any one time. | This structure has the shortcoming of data duplicates of content and unnecessary congestion in the local network in managing the mesh network at the local level. The maintenance of the mesh network is also more expensive as it needs constant updates about the node structure and availability and requires large memory to maintain the mesh information on each and every mesh node. |
| 15. | [21] | Distributed Algorithm for HARP | This structure optimizes the routing between the source node and the receiving nodes in such a manner that the total latency is reduced. This results in the better QoS in terms of reducing the maximum delay between the source and the clients. | The node dynamics has not been considered in constructing the tree and hence node churn would result in broken trees affecting the downstream nodes. |

### III. CONCLUSION

In this work, the author has taken a critical look at the literature on multicasting over overlay networks. Multicasting is one of the most promising applications over the Internet as it requires relatively lower overhead to serve a large number of clients compared to the traditional unicast or broadcast applications. Under the most optimum conditions, multicast systems will have only one data packet in any part of the network irrespective of how many clients are served downstream. Also, the multicast server will transmit only one data packet irrespective of how many clients receive the copies of such packets. In traditional multicast networks, multicast routers placed at strategic locations duplicate the packets depending on the requests they receive. Deploying multicast routers throughout the Internet is not practical due to the amount of investment required for such an operation. Hence researchers have recently focused their attention on using overlay networks to realize the goal of implementing multicast applications in the Internet. By implementing multicasting over application layer overlay networks, the function of duplicating packets has been moved from the network layer (Internet Protocol layer) to application layer.

One of the most challenging tasks in overlay networks in the management of node dynamics. In overlay networks, nodes can join and leave the network at their will. In traditional multicast networks deployed using multicast routers, node dynamics is not a major concern as the infrastructure is considered to be stable and only the leaf or end nodes join and leave the network. The churn of end nodes will not affect any other client node as they are not dependant on each other.

Using overlay networks for multicasting presents a new challenge as the end nodes are required to play a dual role of clients as well as forwarding agents to other client nodes downstream. Node dynamics will have different effects on the end user applications depending on the type of application. Real time applications will be more affected by node dynamics than non-real time applications due to the disruptions resulting from such dynamics.

In this work, media streaming has been selected as the application to be run over the overlay network for the purpose of testing the quality of the overlay networks. Media streaming has been selected as the application due to its popularity and the stringent QoS requirements that are required to be met for successful deployment of such applications in the Internet. Media streaming requires non disrupted flow of packets from the source to destination. Hence managing node dynamics is very important for successful implementation of these applications on the overlay networks.

In this paper, the author presents a critical review of systems, algorithms and mechanism proposed in the recent literature. Special attention has been paid to the advantages and disadvantages of these proposed systems with respect to managing node dynamics. The paper looks at each proposal critically on the mechanism proposed, their strengths and weaknesses. Finally, the results of the analysis have been presented in a table for easy reference.

AUTHOR'S PROFILE

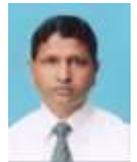

Mohamed Fazil Mohamed Firdhous is a senior lecturer attached to the Faculty of Information Technology of the University of Moratuwa, Sri Lanka. He received his BSc Eng., MSc and MBA degrees from the University of Moratuwa, Sri Lanka, Nanyang Technological University, Singapore and University of Colombo Sri Lanka respectively. In addition to his academic qualifications, he is a Chartered Engineer and a Corporate Member of the Institution of Engineers, Sri Lanka, the Institution of Engineering and Technology, United Kingdom and the International Association of Engineers. Mohamed Firdhous has several years of industry, academic and research experience in Sri Lanka, Singapore and the United States of America.